\documentclass[twocolumn,showpacs,preprintnumbers,amsmath,amssymb]{revtex4}
\usepackage[dvips]{graphicx}
\usepackage{dcolumn}
\usepackage{bm}

\begin{document}

\title
{Antiferromagnetic Order oriented by Fulde-Ferrell-Larkin-Ovchinnikov Superconducting Order}

\author{Yuhki Hatakeyama and Ryusuke Ikeda}

\affiliation{%
Department of Physics, Kyoto University, Kyoto 606-8502, Japan
}

\date{\today}

\begin{abstract}
Resolving the high-field superconducting phase (HFSP), often called as the Q-phase, and the antiferromagnetic or spin-density-wave (SDW) order appearing in the phase remains a crucial issue on the superconductor CeCoIn$_5$. It is shown that a switching of the SDW domain due to a tiny change of the magnetic field direction in HFSP, reported and interpreted as an evidence of the presence of a $\pi$-triplet pairing inducing the SDW order [S. Gerber et al., Nature Physics {\bf 10}, 126 (2014)], can be explained with no triplet pairing component if the $d$-wave superconducting order in HFSP includes the Fulde-Ferrell-Larkin-Ovchinnikov (FFLO) modulation parallel to the field. This result corroborates the picture that HFSP of CeCoIn$_5$ and the SDW order found only in the phase are consequences of the strong paramagnetic pair-breaking in this $d$-wave superconductor. 
\end{abstract}



\maketitle

\section{Introduction}

Unique superconducting (SC) properties in the high field region of the quasi two-dimensional (2D) $d$-wave superconductor CeCoIn$_5$ and, in particular, the presence of its additional high field SC phase (HFSP) in the in-plane field configuration have attracted much interest so far. In 2003, this HFSP has been discovered \cite{Bianchi} and, based on the fact that this material shows an unusually strong paramagnetic pair-breaking (PPB), has been identified with a realization of the Fulde-Ferrell-Larkin-Ovchinnikov (FFLO) state \cite{LOFF} as a vortex phase \cite{AI}. After that, an antiferromagnetic or spin-density-wave (SDW) order has been detected in HFSP in a neutron-scattering measurement \cite{Kenzel}. Different models have been proposed to explain why this SDW order occurs {\it only} in this high field region of the $d_{x^2-y^2}$-wave SC phase. Some of them have found its origin in the strong PPB \cite{IHA,Ilya}, while the others have ascribed its origin to other aspects such as the vortex lattice structure \cite{Machida}, FFLO modulation \cite{Yanase}, and an additional $\pi$-triplet order \cite{Aperis}. 

Recently, a neutron scattering measurement has been reported \cite{Gerber} which detects a sudden change of the direction of the SDW modulation occurring when the magnetic field is rotated within the $a$-$b$ plane through a crystal main axis [100]. In the field precisely parallel to [100], the incommensurate part ${\bf q}$ of the SDW modulation vector can take either of two degenerate directions parallel to the gap nodes of the $d_{x^2-y^2}$-pairing function. The experiment indicates that even a tiny deviation of the field direction from [100] to [110] ([1${\overline 1}$0]) lifts this degeneracy and results in a {\it discontinuous} rotation of ${\bf q}$ to the [1${\overline 1}$0] ([110]) direction. The authors have argued \cite{Gerber} that the scenario \cite{Aperis} requiring the presence of the triplet pairing in HFSP, which induces the SDW order in the $d_{x^2-y^2}$-wave pairing state, is promising and that the pictures ascribing the origin of the SDW {\it only} to some spatial modulation of the SC order parameter are not relevant to this phenomenon. 

Previously, the present authors have proposed the picture \cite{IHA} that HFSP of CeCoIn$_5$ is a coexistent phase of two orders induced by strong PPB, i.e., the SDW order created by an interplay between the $d_{x^2-y^2}$-wave SC pairing and PPB and the FFLO SC order with a spatial modulation parallel to the 
field. The fact that this HFSP is extremely sensitive \cite{Tokiwa} to the purity of the material has been previously interpreted as an evidence of the presence of a spatial modulation parallel to the applied field \cite{RI}. Further, NMR data have clarified a field-dependence of the quasiparticle weight in HFSP which is consistent only with the scenario \cite{IHA,Yanase} invoking the presence of nodal planes perpendicular to the field \cite{Kumagai}. Therefore, it should be clarified whether the neutron data \cite{Gerber} is consistent or not with this FFLO picture. 

In this paper, we theoretically examine a sudden switching \cite{Gerber} of the magnetic domain upon the in-plane field rotation in HFSP of the superconductor CeCoIn$_5$. First, we point out that such a switching of the SDW modulation direction does not occur in any state with a spatially uniform SC order parameter \cite{Ilya,Machida} and show in details that this phenomenon is explained within the picture of the PPB-induced SDW order and without assuming the $\pi$-triplet order if HFSP includes the FFLO spatial modulation parallel to the field \cite{IHA}. This result implies that the strong PPB is the main origin of the presence of HFSP and the SDW order there. 

In sec.II, the model and the procedure of our calculation are sketched, and the main numerical calculation results are presented in sec.III. Summary and comments are given in sec.IV, and the details of the theoretical calculation used here are explained in Appendix. 

\section{Model}

\begin{figure}[t]
\scalebox{0.57}[0.57]{\includegraphics{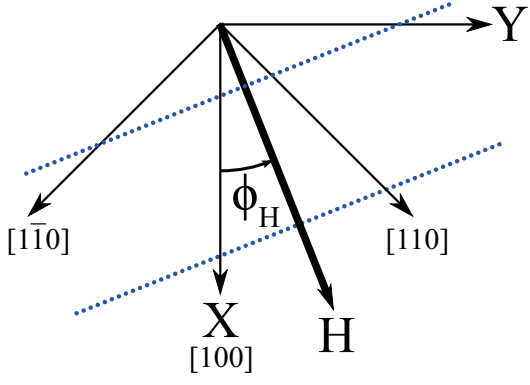}}
\caption{Configuration in $a$-$b$ ($X$-$Y$) plane. The tilt angle $\phi_H$ of the applied field ${\bf H}$ is measured from [100]. The dotted (blue) lines indicate possible FFLO nodal planes perpendicular to ${\bf H} \parallel {\hat x}$.}
\label{fig.1}
\end{figure}

The starting model of our analysis is essentially the same as that used previously \cite{IHA,Hatake} and is an electronic Hamiltonian ${\cal H}={\cal H}_{\rm kin}+{\cal H}_{\rm int}$ of a quasi 2D material. The interaction term ${\cal H}_{\rm int}$ associated with the SC and SDW orders will be treated in the mean field approximation. In zero field, the Hamiltonian is represented in the form 
\begin{eqnarray}
{\cal H}_{\rm kin} &=& \sum_{\bf k} \sum_{\sigma=\pm 1} c_{{\bf k},\sigma}^\dagger \varepsilon({\bf k}) c_{{\bf k},\sigma}, \nonumber \\
{\cal H}_{\rm int} &=& \sum_{\bf p} \biggl[ \frac{1}{U} {\bf m}^*_{{\bf Q},j}({\bf p}) {\bf m}_{{\bf Q},j}({\bf p}) + \frac{1}{g} |\Delta_{\bf p}|^2 \nonumber \\ 
&-& \biggl( {\bf m}^*_{{\bf Q},j}({\bf p}) \sum_{\bf k} {\hat c}^\dagger_{{\bf k}-{\bf p}, \alpha} (\sigma_j)_{\alpha,\beta} {\hat c}_{{\bf k}+{\bf Q}, \beta} 
+ {\rm h.c.} \biggr)
\nonumber \\
&-& \biggl( \Delta^*_{\bf p} \sum_{\bf k} w_{\bf k} \, {\hat c}_{-{\bf k}+{\bf p}/2, \uparrow} {\hat c}_{{\bf k}+{\bf p}/2, \downarrow} 
+ {\rm h.c.} \biggr) \biggr],
\label{model}
\end{eqnarray}
with the dispersion $\varepsilon({\bf k}) = \xi({\bf k}_\perp) - J {\rm cos}(k_ZD)$, where $J$ is an interlayer coupling constant, $D$ is the spacing between the neighboring layers which are parallel to the $a$-$b$ plane, and ${\bf m}_{\bf Q}({\bf p})$ and $\Delta_{\bf p}$ are Fourier components of the SDW and SC order parameters, respectively. Hereafter, the crystal coordinate system $a$-$b$-$c$ will be denoted as $X$-$Y$-$Z$, and a vector ${\bf s}_\perp$ implies a 2D vector perpendicular to ${\hat Z}$. Further, as in the situation in CeCoIn$_5$ in high fields \cite{Kenzel}, ${\bf m}({\bf r})$ is assumed to have only the $c$-axis component, i.e., ${\bf m}=m {\hat Z}$. For a while, the Fermi surface is assumed to be isotropic in the $X$-$Y$ plane, and the in-plane anisotropy will be included later through DOS. Further, $w_{\bf k}=-w_{{\bf k}+{\bf Q}_0}$ is the normalized pairing function with the $d_{x^2-y^2}$-wave symmetry. In a nonzero field ${\bf H} = H {\hat x}$, the Zeeman energy $I \equiv 1.76 \alpha_{{\rm M},c} T_{c0} H/H_{{\rm orb},c}$ needs to be included by shifting $\xi({\bf k}_\perp)$ to $\xi({\bf k}_\perp) + I \sigma$, while the orbital pair-breaking is simply included in terms of the vector-potential ${\bf A}$ by replacing $\xi(-{\rm i}\nabla_\perp)$ with $\xi(-{\rm i}\nabla_\perp + e{\bf A})$, where $T_{c0}$ is the transition temperature in $H=0$, $H_{{\rm orb},c}$ is the orbital limiting field at $T=0$ in ${\bf H} \parallel c$ case, and the constant $\alpha_{{\rm M},c}$ measures the PPB strength \cite{Hatake}. As sketched in Fig.1, the rotated coordinates $x=X {\rm cos}\phi_H + Y {\rm sin}\phi_H$, $y=Y {\rm cos}\phi_H - X {\rm sin}\phi_H$, and $Z=z$ are defined. 

\begin{figure}[t]
\scalebox{0.3}[0.3]{\includegraphics{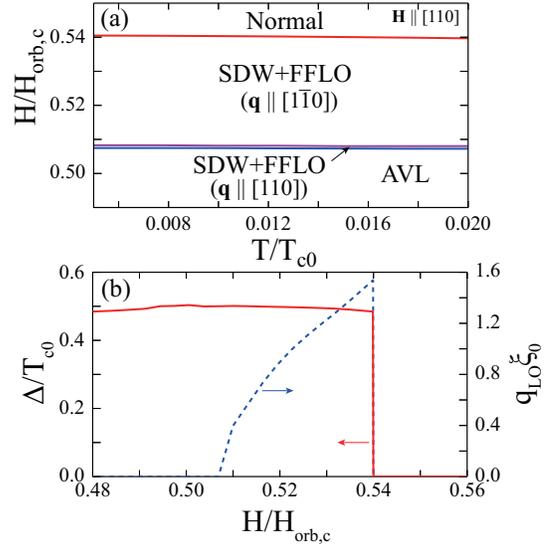}}
\caption{(a) Example of the low $T$ and high $H$ phase diagram in ${\bf H} \parallel$ [110] obtained in terms of the parameters $\gamma=2.77$, $\alpha_{{\rm M},c}=2.5$, $\delta_{\rm IC}=3.0$, and $T_N=0.815 T_{c0}$. The HFSP sandwiched between the normal and the Abrikosov vortex lattice (AVL) phases has ${\bf q} \parallel$ [1${\overline 1}$0] $\perp {\bf H}$ except in quite a narrow range close to the second order transition to AVL. See the text for details. (b) Field dependences of the resulting energy gap $|\Delta|$ and the wave number $q_{\rm LO}$, normalized by the coherence length $\xi_0=\hbar v_{\rm F}/(2 \pi T_{c0})$, of the FFLO modulation at $T=0.02 T_{c0}$ in (a). }
\label{fig.2}
\end{figure}

The SDW ${\bf Q}$-vector is the sum of a commensurate component ${\bf Q}_0$ (one of ($\pm\pi$, $\pm\pi$, $\pm\pi$)) and an incommensurate one 
\begin{equation}
{\bf q} = |{\bf q}|({\hat X} {\rm cos}\phi_{\bf q}+{\hat Y}{\rm sin}\phi_{\bf q}). 
\end{equation}
In the ensuing expression of the free energy, this incommensurate part ${\bf q}$ will appear in the form $\delta({\bf k})=\varepsilon({\bf k}+{\bf Q}_0)+\varepsilon({\bf k})-{\bf v}_{\bf k}\cdot{\bf q}$ and be determined by minimizing the free energy, where ${\bf v}_{\bf k}=\partial \varepsilon({\bf k})/\partial {\bf k}$. However, the SDW order parameter can have other spatial modulations to lower the energy through a coupling to the spatially varying SC order parameter $\Delta$. In eq.(\ref{model}), possible spatial modulations of $\Delta$ and the above-mentioned additional modulation of the SDW order parameter ${\bf m}$ are represented by their ${\bf p}$-dependences. 

The mean field free energy density $F$ is derived following a familiar route \cite{IHA,Ilya,Hatake} and 
consists of three terms, i.e. $F=F_{\rm SC}+F_{\rm SDW}+F_{\rm cl}$. Their details can be seen in Appendix. The SC term $F_{\rm SC}$ is the familiar GL expansion in $\Delta$ kept up to the O($|\Delta|^6$) term. In truncating the expansion to the sixth-order term, we have verified that the coefficient of this term is positive. As in previous works, we follow the picture \cite{AI,IHA,Yanase,Hatake,RI} that HFSP occurs due to the formation of a FFLO spatial modulation of $\Delta$ parallel to ${\bf H} \parallel {\hat x}$. In this case, the SC order parameter $\Delta({\bf r})$ in the coordinate representation takes the form 
\begin{equation}
\Delta({\bf r})=\sqrt{2} \Delta_0(y,z) {\rm cos}(q_{\rm LO}x). 
\end{equation}
This solution has nodal planes which are parallel to the $y-z$ plane and periodic in $x$. 
Regarding the vortex lattice structure expressed by $\Delta_0(y,z)$, the Abrikosov lattice solution in the lowest Landau level under the in-plane field $H {\hat x}$ in a system with an uniaxial anisotropy will be used. The anisotropy is measured by the parameter $\gamma$ ($ > 1$) which, roughly speaking, corresponds to the ratio of the in-plane and out-of-plane coherence lengths and is determined by the velocity ${\bf v}_{\bf k}$ and the gap function's magnitude $|w_{\bf k}|$ (see Appendix). Since the $H_{c2}$-transition is discontinuous reflecting the strong PPB \cite{AI} (see Fig.2), the magnitude $|\Delta|$ of the SC order parameter is rigid anywhere below $H_{c2}$ so that the FFLO spatial order is stabilized irrespective of the appearance of the SDW order. For this reason, it will be assumed that the presence of the SDW order does not affect the details of the FFLO order. 

Below, we focus on other free energy terms associated with the SDW order parameter $m_{\bf Q}({\bf r})$. 
As already mentioned, $m_{\bf Q}$ should have an additional spatial modulation induced by the FFLO modulation of $\Delta$ (see eq.(3)) with the wave vector ${\bf q}_{\rm LO} \parallel {\bf H}$. According to the conventional treatment \cite{Maki} on the metallic SDW ordering,  the SDW free energy density $F_{\rm SDW}$ unaccompanied by the SC order parameter is given by
\begin{widetext} 
\begin{equation}
\frac{F_{\rm SDW}}{N(0)} = \biggl\langle {\rm ln}\biggl(\frac{T}{T_N} \biggr) + {\rm Re}\biggl[\psi\biggl(\frac{1}{2} + {\rm i}\frac{\delta({\bf k})}{4 \pi T} \biggr) - \psi\biggl(\frac{1}{2}\biggr) \biggr] \biggr\rangle_{\hat {\bf k}} \langle |m|^2 \rangle_{\rm s} 
- \biggl\langle {\rm Re}\biggl[\psi^{(2)}\biggl(\frac{1}{2}+{\rm i}\frac{\delta({\bf k})}{4 \pi T} \biggr) \biggr] \biggr\rangle_{\hat {\bf k}} \frac{\langle |m|^4 \rangle_{\rm s}}{(4 \pi T)^2}, \label{eq:FSDW}
\end{equation}
\end{widetext}
where $\psi(x)$ and $\psi^{(2)}(x)$ are the digamma function and its second derivative, respectively, and $T_N$ is a Neel temperature in the commensurate limit. We choose parameter values for which the coefficient of the $|m|^2$ term in $F_{\rm SDW}$ remains positive at any temperature. That is, as in the situation of CeCoIn$_5$ in ${\bf H} \perp c$, we focus on the case with no SDW order in the normal phase. We have verified that, for those parameter values, the coefficient of the $|m|^4$ term in $F_{\rm SDW}$ is positive. 

The SDW order is induced by the following coupling term 
$F_{\rm cl}$ between the two orders \cite{IHA}. Up to the lowest order in $|\Delta|^2$, it takes the form 
\begin{eqnarray}
F_{\rm cl}\!\!\!&=&\!\!\! \int_0^\infty d\rho \int_0^\infty d\Lambda \frac{4 \pi T N(0) \langle |\Delta|^2 \rangle_s \langle |m|^2 \rangle_s}{{\rm sinh}(2 \pi T (\rho+2\Lambda))} \int_{-\infty}^\infty d\tau \nonumber \\
&\times&\!\!\! \langle |w_{\bf k}|^2 [K^{(n)}(\Lambda,\tau,\rho; {\hat {\bf k}}) + K^{(an)}(\Lambda, \tau, \rho; {\hat {\bf k}})] \rangle_{\hat {\bf k}}, \label{eq:Fcl}
\end{eqnarray}
where 
\begin{widetext}
\begin{eqnarray}
K^{(n)} &=& [{\rm cos}(I(2\Lambda-\tau)) {\rm cos}(\delta({\bf k})(\Lambda+\rho+\tau/2)) e^{-|\eta_{\bf k}|^2(\Lambda-\tau/2)^2/2} 
+ {\rm cos}(4I\Lambda) {\rm cos}(\delta({\bf k})\rho) e^{-2|\eta_{\bf k}|^2\Lambda^2}  ](2 {\rm cos}(2q_{\rm LO}v_{{\bf k},x}\Lambda)  \nonumber \\
&\times& {\rm cos}(q_{\rm LO}v_{{\bf k},x}\rho) 
+ {\rm cos}(q_{\rm LO}v_{{\bf k},x}(\tau+\rho))), \nonumber \\
K^{(an)} &=& - {\rm cos}(2I\tau) {\rm cos}(\delta({\bf k})\rho) e^{-|\eta_{\bf k}|^2 \tau^2/2} [ 2{\rm cos}(q_{\rm LO}v_{{\bf k},x}\tau) 
{\rm cos}(q_{\rm LO}v_{{\bf k},x}\rho) + {\rm cos}(q_{\rm LO}v_{{\bf k},x}(2\Lambda + \rho)) ], \label{eq:K}
\end{eqnarray}
\end{widetext}
and $|\eta_{\bf k}|^2=|e|H(v_{{\bf k},y}^2 + \gamma^2 v_{{\bf k},z}^2 )/\gamma$ which depends on the ${\bf H}$-direction, i.e., on $\phi_H$. Derivation of the above expressions, presented in Appendix, is a simple extension of the GL approach in the previous works \cite{IHA,Hatake} to the case with the FFLO modulation. Here, we have assumed the FFLO modulation $m_{\bf Q}(x) \propto {\rm cos}(q_{\rm LO}x)$ with the same phase as eq.(3) because, up to the lowest order in $|\Delta|^2$, the SDW order favors the region in real space with a nonvanishing $\Delta$ rather than that with $\Delta=0$ \cite{Hatake}. 

The sign of $K^{(n)}$ and $K^{(an)}$ is determined by trigonometric factors of different origins, the PPB effect due to the Zeeman energy, the ${\bf q}$-direction reflected in $\delta({\bf k})$, and effects of a nonvanishing $q_{\rm LO}$. On the other hand, the magnitudes of $K^{(n)}$ and $K^{(an)}$ are affected by the exponential factor reflecting the presence of vortices. Roughly speaking, this exponential factor selects the component nearly parallel to ${\bf H}$ of ${\bf k}$ on the Fermi surface. The SDW order is present when 
\begin{equation}
\chi(\phi_{\bf q}) \equiv \frac{F_{\rm SDW}+F_{\rm cl}}{\langle |m|^2 \rangle_s} \biggr|_{m=0}
\end{equation} 
is negative. The direction of ${\bf q}$, which is the main focus in this paper, is determined through minimizing the free energy. 

\section{Results}

\begin{figure}[t]
\scalebox{0.3}[0.3]{\includegraphics{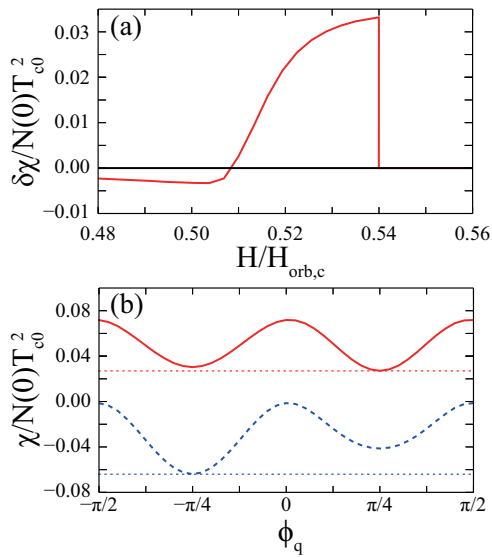}}
\caption{(a) Field dependence of $\delta \chi \equiv \chi(\phi_{\bf q}=\pi/4) - \chi(\phi_{\bf q}=-\pi/4)$ in ${\bf H} \parallel$ [110] following from the same set of parameters as in Fig.2(b).  (b) Corresponding $\chi(\phi_{\bf q})$ curves in $H=0.5 H_{{\rm orb},c}$ (solid red curve) and in $H=0.52 H_{{\rm orb},c}$ (blue dashed one).}
\label{fig.3}
\end{figure}

Among the obtained results in the present work, let us first discuss the ${\bf q}$-orientation in ${\bf H} \parallel$ [100], i.e., $\phi_H=0$, case. In this case, the expressions are symmetric in the sign of $k_y$, and thus,  the two configurations symmetric with respect to [100] are degenerate in energy with each other. Further, the $H_{c2}$-values in the present system with strong PPB is not so large that the two-fold symmetry due to the vortices in the $X$-$Y$ plane is a weaker effect compared with the four-fold symmetry of the pairing function $|w_{\bf k}|$. Consequently, the free energy density $F$ has its extreme values around $\phi_{\bf q} = \pm \pi/4$, while the curvature $\partial^2 F/\partial \phi_{\bf q}^2$ depends on the magnitude of the incommensurability $|\delta_{\rm IC}| = |\varepsilon({\bf k}) + \varepsilon({\bf k}+{\bf Q}_0)|/T_{c0}$. Typically, for larger $|\delta_{\rm IC}|$ ($> 1$), $F$ is minimized around $\phi_{\bf q}=\pm \pi/4$ (see Fig.3). We note that the PPB-induced SDW order tends to be enhanced with increasing $|\delta_{\rm IC}|$ \cite{IHA}. In our calculation results which are shown hereafter, the value $|\delta_{\rm IC}|=3$ has been used. 

Once $\phi_H$ becomes nonzero, however, the degeneracy is lifted by the presence of the vortices and the FFLO modulation. Interestingly, these two effects favor {\it different} orientations of ${\bf q}$ from each other. To see this, let us first focus on the $q_{\rm LO}=0$ case, i.e., the ordinary vortex lattice with no FFLO modulation, by assuming $\phi_H > 0$. As Fig.3(b) shows, the free energy in lower fields than HFSP, i.e., $H < 0.508 H_{\rm orb,c}$, is lower when $\phi_{\bf q} > 0$, implying the tendency that ${\bf q}$ is oriented along the vortex axis parallel to ${\bf H}$. This feature has also been verified elsewhere \cite{Machida}. If HFSP is merely a coexistent phase of 
the ordinary vortex lattice with the $d_{x^2-y^2}$-wave SC pairing and a SDW order \cite{Ilya,Machida}, $\phi_{\bf q}$ would has the same sign as that of $\phi_H$, in contrast to the experimental observation \cite{Gerber}. 

Therefore, HFSP must have a different factor for changing the sign of $\phi_{\bf q}$. According to the original proposal on HFSP of CeCoIn$_5$ \cite{Bianchi,AI}, we next examine the corresponding results in the case with the FFLO 
modulation. 
In eq.(5), the sign of $\phi_{\bf q}$ minimizing the free energy 
is determined by 
keeping the sign of the product of two kinds of trigonometric factors, the factor including ${\bf q}$ and that including $q_{\rm LO}$, unchanged: For instance, in $K^{(an)}$, the sign of ${\rm cos}(\delta({\bf k})\rho)$ is reversed by a large change of ${\bf q}$-direction, 
because the dominant ${\bf k}$-direction is, as already mentioned, limited 
by the orbital pair-breaking, and this sign reversal is compensated rather by sign changes of other trigonometric factors including $q_{\rm LO}$. 

In Fig.4, the resulting $\phi_H$-dependence of $\phi_{\bf q}$ is shown as a solid curve. 
The use of eq.(3) with a nonzero $q_{\rm LO}$ leads to the result that the free energy is lowered in the configuration with $\phi_{\bf q} \phi_H < 0$, suggesting that the orientation ${\bf q} \perp {\bf H}$ is favored, in contrast to that in the $q_{\rm LO}=0$ case. In a narrow region in the close vicinity of the second order transition entering HFSP where $|q_{\rm LO}| \xi_0$ is small ($< 0.2$), the configuration $\phi_{\bf q} \phi_H > 0$ is realized, as in the low field vortex lattice (see Fig.2). For larger $q_{\rm LO} \xi_0$ of order unity, however, the FFLO modulation acts on the ${\bf q}$ orientation more strongly than the anisotropy due to the vortices, and the configuration $\phi_{\bf q} \phi_H < 0$ results in, although $\phi_{\bf q}$ favors values,  more or less, close to $\pm \pi/4$ due to the four-fold symmetry of the gap function $w_{\bf k}$. 
Physically, it implies that ${\bf q}$ tends to be oriented along the FFLO nodal planes. This "pinning" of ${\bf q}$ to the nodal planes seems to be the origin of the quick approach of the ${\bf q}$-vector to [1${\overline 1}$0] as $\phi_H$ is slightly increased from zero. In fact, it is clear from Fig.1 that, according to the above-mentioned pinning effect, the tilt of the nodal plane due to a slight and positive (negative) $\phi_H$ favors $\phi_{\bf q} =-\pi/4$ ($+\pi/4$). Further, since the effect of the FFLO modulation on the ${\bf q}$-orientation is much bigger for the PPB strength used here (see Fig.3) than that of the in-plane anisotropy due to the vortices, a change of $\phi_H$ at ${\bf H} \parallel$ [100] with $\phi_{\bf q} \phi_H > 0$, expected in the ordinary vortex lattice, does not occur in this case. In addition, the feature seen in the solid curve of Fig.4 that $|\phi_{\bf q}| > \pi/4$ for smaller $|\phi_H|$ values can also be understood from Fig.1 by taking account of this pinning of ${\bf q}$ to the nodal planes. 

The dashed curve in Fig.4 shows the corresponding $\phi_{\bf q}$ v.s. $\phi_H$ curve obtained in a more realistic case with a larger DOS along [110]. In the present approach, the anisotropy on DOS is incorporated with the replacement of the normal DOS on the Fermi surface 
$N(0) \to N(0)/(1 + \beta {\rm cos}(4 \phi_{\bf k}))$ with $\beta > 0$ \cite{Nakai}, where $\phi_{\bf k}={\rm tan}^{-1}(k_Y/k_X)$. Since this four-fold anisotropy merely suppresses the deviation, seen in the solid curve, of $|\phi_{\bf q}|$ from $\pi/4$ in the small $|\phi_H|$ range, it seems that the solid curve following from the isotropic Fermi surface includes all of essential contents of the $\phi_H$-dependent ${\bf q}$-orientation. 

\begin{figure}[t]
\scalebox{0.35}[0.35]{\includegraphics{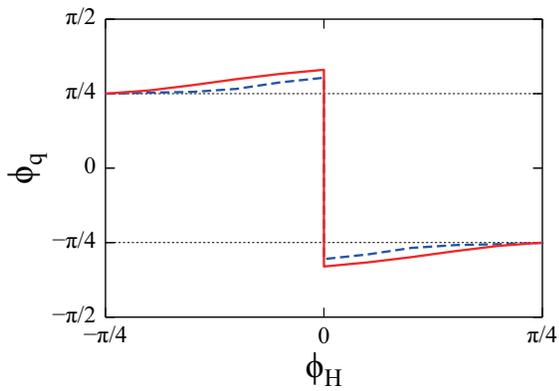}}
\caption{ $\phi_H$ v.s. $\phi_{\bf q}$ curves at $H=0.51 H_{{\rm orb},c}$ and $T=0.01 T_{c0}$ in the case (solid curve) where the in-plane Fermi surface is isotropic and the case (dashed one) with a Fermi surface anisotropy incorporated through the replacement of DOS with $\beta=0.1$ (see the text). The used $T_N/T_{c0}$-value is $0.86$ in the solid curve and $0.91$ in the dashed one, respectively, and other parameters are $\gamma=2.12$, $\delta_{\rm IC}=3.0$, and $\alpha_{{\rm M},c}=2.5$ in both curves. }
\label{fig.4}
\end{figure}

\section{Summary and Comments}

As shown in the preceding section, the sudden switching of the magnetic domain due to a slight rotation of ${\bf H}$ around the [100] direction seen in HFSP of the $d_{x^2-y^2}$-wave superconductor CeCoIn$_5$ \cite{Gerber} can be explained, based on the original picture \cite{Bianchi,AI} that HFSP is a FFLO superconducting phase, as an event stemming from a {\it pinning} of the SDW ${\bf Q}$-vector to the FFLO nodal planes. This FFLO picture of HFSP has been supported previously through the NMR \cite{Kumagai} and doping \cite{Tokiwa} 
experiments and a related theoretical study 
\cite{RI}. It should be stressed here that the origin of the SDW order is not the FFLO modulation of $\Delta$ but consists in an interplay between the PPB effect and the $d_{x^2-y^2}$-pairing symmetry \cite{IHA}. 
On the other hand, it has been argued in Ref.\cite{Gerber} that the observed switching of the magnetic domain is an evidence of the presence of a $\pi$-triplet order in HFSP. Justification of this phenomenology \cite{Gerber} would need to be accompanied by a firm microscopic basis for the presence of such a rare pairing state. In fact, as pointed out previously \cite{RI}, it is difficult to explain the strong doping effect \cite{Tokiwa} of HFSP based only on the presence of the $\pi$-triplet order. We also note that observed changes of HFSP on tilting the applied field from the $a$-$b$ plane have also been explained based on this FFLO-based theory \cite{Hosoya}.

One of the authors (Y.H.) thanks Y. Yanase for valuable discussions. Y.H. is supported by JSPS Research Fellowship for Young Scientists, and 
the research of R.I. was supported by Grant-in-Aid for Scientific Research [No. 25400368] from MEXT, Japan. 

\vspace{2mm}

\appendix*
\section{}
Here, the detailed derivation of the mean field free energy density $F$ is presented.
In this derivation, we use the perturbative approach adopted in Ref.\cite{IHA, Hatake} by refining it in a form incorporating the coupling between the FFLO modulation of a SC order parameter and the SDW ${\bf q}$-vector. 

We consider the GL expansion of the mean field free energy density $F= F_{\mathrm{SC}}+F_{\mathrm{SDW}}+F_{\mathrm{cl}}$ up to $O(|\Delta|^6)$, $O(|m|^4)$, and $O(|\Delta|^2|m|^2)$ terms:
\begin{align}
	F_{\mathrm{SC}}=& F_{\Delta}^{\mathrm{(2)}}+F_{\Delta}^{\mathrm{(4)}}+F_{\Delta}^{\mathrm{(6)}}, \label{eqApp:FSC}\\
	F_{\mathrm{SDW}}=& F_{m}^{\mathrm{(2)}}+F_{m}^{\mathrm{(4)}}, \label{eqApp:FSDW} \\
	F_{\mathrm{cl}}=& F_{\Delta m}^{\mathrm{(2,2)}}. \label{eqApp:Fcl}
\end{align}
Here, $F_{\Delta}^{\mathrm{(N)}}$ ($N=2,4,6$) and $F_{m}^{\mathrm{(M)}}$ ($M=2,4$) denote the expansion terms proportional to $|\Delta|^N$ and $|m|^M$, respectively, and $F_{\Delta m}^{\mathrm{(2,2)}}$ represents the coupling term between SC and SDW orders, which is proportional to $|\Delta|^2|m|^2$.

In order to incorporate the orbital pair-breaking effect, the quasi-classical approximation of the Green function $\mathcal{G}_{\omega_n,\sigma}(\bm{r},\bm{r^\prime})$ is employed:
\begin{equation} 
	\mathcal{G}_{\omega_n,\sigma}(\bm{r},\bm{r^\prime})
	\simeq \mathcal{G}_{\omega_n,\sigma}(\bm{r-r^\prime})|_{\bm{A}=0} \times e^{ ie \int^{\bm{r}}_{\bm{r^\prime}} \bm{A}(\bm{s}) \cdot d\bm{s}} ,
\end{equation}
where $\omega_n$ is the fermion Matsubara frequency, and 
\begin{align}
	\mathcal{G}_{\omega_n,\sigma}(\bm{r-r'})|_{\bm{A}=0} &=\sum_{\bm{k}} 	\mathcal{G}_{\omega_n,\sigma}(\bm{k})e^{i\bm{k}\cdot(\bm{r-r'})}, \\
	\mathcal{G}_{\omega_n,\sigma}(\bm{k}) &= \frac{1}{i\omega_n - \varepsilon(\bm{k}) - I\sigma} .
\end{align}

Using the formula \cite{Wert}
\begin{equation}
e^{ 2ie \int^{\bm{r}_1}_{\bm{r}} \bm{A}(\bm{s}) \cdot d\bm{s} } \Delta(\bm{r}_1)
	= e^{i(\bm{r}-\bm{r}_1) \cdot \bm{\Pi} } \Delta(\bm{r}) ,
\end{equation}
where $\bm{\Pi} = -i\nabla + 2e \bm{A}$,
$F_{\Delta}^{\mathrm{(N)}}$ ($N=2,4,6$) is straightforwardly calculated in the form 
\begin{equation} 
	F_{\Delta}^{\mathrm{(2)}} = \left\langle \Delta^*(\bm{r}) \left[ \frac{1}{|g|} - K_{\Delta}^{\mathrm{(2)}}(\bm{\Pi}) \right] \Delta(\bm{r}) \right\rangle_{\mathrm{s}} , \label{eqApp:fD2}
\end{equation}
\begin{equation} 
	F_{\Delta}^{\mathrm{(4)}} = \left\langle  K_{\Delta}^{\mathrm{(4)}}(\bm{\Pi}_i) \Delta^*(\bm{r}_1) \Delta(\bm{r}_2) \Delta^*(\bm{r}_3) \Delta(\bm{r}_4) |_{\bm{r}_i\to\bm{r}} \right\rangle_{\mathrm{s}} , \label{eqApp:fD4}
\end{equation}
\begin{align} 
	F_{\Delta}^{\mathrm{(6)}} =& \Big\langle K_{\Delta}^{\mathrm{(6)}}(\bm{\Pi}_i) \Delta^*(\bm{r}_1) \Delta(\bm{r}_2) \Delta^*(\bm{r}_3) 
\nonumber \\  & \times 
	\Delta(\bm{r}_4) \Delta^*(\bm{r}_5) \Delta(\bm{r}_6)|_{\bm{r}_i\to\bm{r}} \Big\rangle_{\mathrm{s}} , \label{eqApp:fD6}
\end{align}
where $\langle \rangle_{\mathrm{s}}$ denotes the spatial average, and
\begin{align} 
	K_{\Delta}^{\mathrm{(2)}}(\bm{\Pi}) &= \frac{T}{2} \sum_{\omega_n,\bm{k},\sigma} |w_{\bm{k}}|^2 \mathcal{G}_{\omega_n,\sigma}(\bm{k}) 
	\mathcal{G}_{-\omega_n,-\sigma}(-\bm{k}+\bm{\Pi}) \nonumber \\
	&= 2\pi T N(0) \int_0^\infty d\rho f(\rho) \left\langle |w_{\bm{k}}|^2 e^{-i\rho \bm{v_k}\cdot\bm{\Pi}} \right\rangle_{\hat{\bm{k}}} , \label{eqApp:KD2}
\end{align}
\begin{widetext}
\begin{align} 
	K_{\Delta}^{\mathrm{(4)}}(\bm{\Pi}_i) =& \frac{T}{4} \sum_{\omega_n,\bm{k},\sigma} |w_{\bm{k}}|^4 \mathcal{G}_{\omega_n,\sigma}(\bm{k}) 
	\mathcal{G}_{-\omega_n,-\sigma}(-\bm{k}+\bm{\Pi^*}_1) 
	\mathcal{G}_{-\omega_n,-\sigma}(-\bm{k}+\bm{\Pi}_2)
	\mathcal{G}_{\omega_n,\sigma}(\bm{k}+\bm{\Pi^*}_3-\bm{\Pi}_2) \nonumber \\
	=& 2\pi T N(0) \int_0^\infty \prod_{i=1}^3 d\rho_i\ f \left( \sum_{i=1}^3\rho_i \right) 
	\left\langle |w_{\bm{k}}|^4
	e^{i\bm{v_k}\cdot(\rho_1\bm{\Pi^*}_1+\rho_2\bm{\Pi}_2+\rho_3\bm{\Pi^*}_3)} \right\rangle_{\hat{\bm{k}}}  
	+  (\bm{\Pi}_2 \leftrightarrow \bm{\Pi}_4) ,\label{eqApp:KD4}
\end{align}
\begin{align} 
	K_{\Delta}^{\mathrm{(6)}}(\bm{\Pi}_i) =& -\frac{T}{6} \sum_{\omega_n,\bm{k},\sigma} |w_{\bm{k}}|^6 \mathcal{G}_{\omega_n,\sigma}(\bm{k}) 
	\mathcal{G}_{-\omega_n,-\sigma}(-\bm{k}+\bm{\Pi^*}_1) 
	\mathcal{G}_{-\omega_n,-\sigma}(-\bm{k}+\bm{\Pi}_6) 
\nonumber \\ & \times
	\mathcal{G}_{\omega_n,\sigma}(\bm{k}-\bm{\Pi^*}_1-\bm{\Pi}_2)  	
	\mathcal{G}_{-\omega_n,-\sigma}(-\bm{k}+\bm{\Pi^*}_1+\bm{\Pi^*}_3-\bm{\Pi}_2)
	\mathcal{G}_{\omega_n,\sigma}(\bm{k}-\bm{\Pi}_6+\bm{\Pi^*}_5) \nonumber \\
	=& - 2\pi T N(0) \int_0^\infty \prod_{i=1}^5 d\rho_i\ f \left( \sum_{i=1}^5\rho_i \right) \left\langle |w_{\bm{k}}|^6
	e^{i\bm{v_k}\cdot(\rho_1\bm{\Pi^*}_1+\rho_2\bm{\Pi}_2+\rho_3\bm{\Pi^*}_3+\rho_4\bm{\Pi}_4+\rho_5\bm{\Pi^*}_5)}\right\rangle_{\hat{\bm{k}}}
\nonumber \\ &
	+(\bm{\Pi^*}_3 \rightarrow \bm{\Pi}_2-\bm{\Pi^*}_3+\bm{\Pi}_4). \label{eqApp:KD6}
\end{align}
Here, $\langle \rangle_{\hat{\bm{k}}}$ represents the k-space average on the Fermi surface, and $f(\rho) = \cos(2I\rho)/\sinh(2\pi T\rho)$.
In order to obtain eqs.(\ref{eqApp:KD2}), (\ref{eqApp:KD4}), and (\ref{eqApp:KD6}), the identity 
$1/\alpha=\int_0^{\infty} d\rho \ e^{-\alpha\rho}$ ($\mathrm{Re}[\alpha]>0$) is used.

Similarly, the expressions for $F_{m}^{\mathrm{(M)}}$ ($M=2,4$) are written as
\begin{align} 
	F_{m}^{\mathrm{(2)}} &= \left\langle \frac{1}{U} - \frac{T}{2} \sum_{\omega_n,\bm{k},\sigma} \mathcal{G}_{\omega_n,\sigma}(\bm{k}) 
	\mathcal{G}_{\omega_n,-\sigma}(\bm{k+Q}) \right\rangle_{\hat {\bf k}} \langle |m(\bm{r})|^2 \rangle_{\mathrm{s}} , \label{eqApp:fm2}
\end{align}
\begin{align} 
	F_{m}^{\mathrm{(4)}} &= \left\langle \frac{T}{2} \sum_{\omega_n,\bm{k},\sigma} 
	\mathcal{G}_{\omega_n,\sigma}(\bm{k}) \mathcal{G}_{\omega_n,-\sigma}(\bm{k+Q})
	 \mathcal{G}_{\omega_n,\sigma}(\bm{k}) \mathcal{G}_{\omega_n,-\sigma}(\bm{k+Q}) \right\rangle_{\hat {\bf k}} \langle |m(\bm{r})|^4 \rangle_{\mathrm{s}} , \label{eqApp:fm4}
\end{align}
where $m(\bm{r})=\sum_{\bm{p}} m_{\bm{Q}}(\bm{p})
\exp(i\bm{p}\cdot\bm{r})$. 
Substituting eqs.(\ref{eqApp:fm2}) and (\ref{eqApp:fm4}) into eq.(\ref{eqApp:FSDW}), and using the expression
$1/U=N(0)[\ln(T/T_N)+\sum_{\omega_n>0}2\pi T/\omega_n]$, we obtain eq.(\ref{eq:FSDW}) in the main text.

Similarly, we can calculate $F_{\Delta m}^{\mathrm{(2,2)}}$ in the form 
\begin{equation} 
	F_{\Delta m}^{\mathrm{(2,2)}} = \left\langle \left[ 2K_{\Delta m,1}(\bm{\Pi}_i,-i\nabla_i) + K_{\Delta m,2}(\bm{\Pi}_i,-i\nabla_i)
	 \right] \Delta^*(\bm{r}_1) \Delta(\bm{r}_2) m^*(\bm{r}_3) m(\bm{r}_4) |_{\bm{r}_i\to\bm{r}} \right\rangle_{\mathrm{s}} , \label{eqApp:fD2m2} 
\end{equation}
where
\begin{align} 
	 K_{\Delta m,1}(\bm{\Pi}_i,-i\nabla_i) &= -T \sum_{\omega_n,\bm{k},\sigma} |w_{\bm{k}}|^2 
	\mathcal{G}_{\omega_n,\sigma}(\bm{k-Q}+i\nabla_3)
	\mathcal{G}_{\omega_n,-\sigma}(\bm{k}) 
	\mathcal{G}_{-\omega_n,\sigma}(-\bm{k}+\bm{\Pi}_2) 
	\mathcal{G}_{\omega_n,-\sigma}(\bm{k}-\bm{\Pi}_2+\bm{\Pi^*}_1)  \nonumber \\
	&= \int_0^\infty \prod_{i=1}^3 d\rho_i \frac{2\pi T N(0)}{\sinh[2\pi T(\sum_{i=0}^3\rho_i)]} \Bigg\langle |w_{\bm{k}}|^2
	\left[ \cos\left(2I(\rho_1+\rho_2)\right) e^{i\delta(\bm{k})\rho_3}
	e^{-i\bm{v_k}\cdot(\rho_1\bm{\Pi^*}_1+\rho_2\bm{\Pi}_2-\rho_3i\nabla_3)}
\nonumber \right. \\ & \left. \quad
	+ \cos\left(2I\rho_2\right) e^{i\delta(\bm{k})(\rho_1+\rho_3)} 
	e^{-i\bm{v_k}\cdot\left ((\rho_1+\rho_2)\bm{\Pi^*}_1-\rho_1\bm{\Pi}_2- (\rho_1+\rho_3)i\nabla_3 \right)} \right] + \mathrm{h.c.} \Bigg\rangle_{\hat{\bm{k}}} , \label{eqApp:K1}
\end{align}
\begin{align} 
	K_{\Delta m,2}(\{\bm{\Pi}_i,-i\nabla_i\}) =& -T \sum_{\omega_n,\bm{k},\sigma} w_{\bm{k}}w^*_{\bm{k+Q}} \mathcal{G}_{\omega_n,\sigma}(\bm{k}+\bm{\Pi}_2)
	\mathcal{G}_{-\omega_n,-\sigma}(\bm{-k}) 
\nonumber \\ & \times
	\mathcal{G}_{-\omega_n,\sigma}(\bm{-k-Q}+i\nabla_3) \mathcal{G}_{\omega_n,-\sigma}(\bm{k+Q}+\bm{\Pi^*}_1-i\nabla_3) \nonumber \\
	=& - \int_0^\infty \prod_{i=1}^3 d\rho_i \frac{2\pi TN(0)}{\sinh[2\pi T(\sum_{i=0}^3\rho_i)]} 
	 \left\langle |w_{\bm{k}}|^2
	\cos\left(2I(\rho_1-\rho_2)\right) e^{i\delta(\bm{k})\rho_3} 
\nonumber \right. \\ & \left. \times
	\left[ e^{i\bm{v_k}\cdot(\rho_1\bm{\Pi^*}_1-\rho_2\bm{\Pi}_2+\rho_3i\nabla_3)}
	 +e^{i\bm{v_k}\cdot\left( (\rho_1+\rho_3)\bm{\Pi^*}_1-(\rho_2+\rho_3)\bm{\Pi}_2-\rho_3i\nabla_3 \right)} \right] 
	+ \mathrm{h.c.} \right\rangle_{\hat{\bm{k}}} . \label{eqApp:K2}
\end{align}

As discussed in the main text, the SC and SDW order parameters in the coordinate representation are given by 
\begin{align}
	\Delta(\bm{r})=& \sqrt{2}\Delta_0(y,z)\cos(q_{\mathrm{LO}}x) , \label{eqApp:Delta} \\
	m(\bm{r})=& \sqrt{2}m\cos(q_{\mathrm{LO}}x) . \label{eqApp:m}
\end{align}
Here, $\Delta_0(y,z)$ is the Abrikosov lattice solution defined in the anisotropic plane :
\begin{align} 
	\Delta_0(y,z) =& \Delta \left( \frac{k^2}{\pi} \right)^{\frac{1}{4}} \sum_{n=-\infty}^\infty \exp \left[i \left( \frac{nk}{r_H\sqrt{\gamma}} z + \frac{\pi}{2} n^2 \right)
	- \frac{1}{2} \left( \frac{\sqrt{\gamma}}{r_H}y + nk \right)^2 \right] ,
\end{align}
where $r_H = (2|eH|)^{-1/2}$, and $\gamma = \sqrt{ \left\langle |w_{\bm{k}}|^2v_{\bm{k},y}^2 \right\rangle_{\hat{\bm{k}}}/\left\langle |w_{\bm{k}}|^2v_{\bm{k},z}^2 \right\rangle_{\hat{\bm{k}}} }$($v_{\bm{k},y}$ and $v_{\bm{k},z}$ are the $y$ and $z$ components of $\bm{v_k}$ in the rotated coordinates, respectively). Further, for simplicity, the square lattice solution with $k = \sqrt{\pi}$ has been adopted. We note that the type of the vortex lattice does not affect our main results even quantitatively. 

Substituting eqs.(\ref{eqApp:Delta}) and (\ref{eqApp:m}) into eqs.(\ref{eqApp:fD2}), (\ref{eqApp:fD4}), (\ref{eqApp:fD6}), and (\ref{eqApp:fD2m2}), and employing the local approximation \cite{AI}, we obtain the expressions for $F_{\Delta}^{\mathrm{(N)}}$ ($N=2,4,6$) and $F_{\Delta m}^{\mathrm{(2,2)}}$ as follows:
\begin{align}
	F_{\Delta}^{\mathrm{(2)}}=& N(0)\left[ \ln \left(\frac{T}{T_{c0}}\right)+ 2\pi T\int_0^\infty d\rho \left\langle |w_{\bm{k}}|^2 \left( \frac{1}{\sinh(2\pi T\rho)}
	- f(\rho)  e^{-|\eta_{\bm{k}}|^2\rho^2/2} \cos\left(q_{\mathrm{LO}}v_{\bm{k},x}\rho\right) \right) \right\rangle_{\hat{\bm{k}}} \right] |\Delta|^2 ,
\end{align}
\begin{align}
	& F_{\Delta}^{\mathrm{(4)}}=  \frac{\pi c_4 T N(0)}{\sqrt{2}} \int_0^\infty \prod_{i=1}^3 d\rho_i\ f \left(\sum_{i=1}^3\rho_i \right) \Bigg\langle |w_{\bm{k}}|^4 
	\exp\left[ -\frac{|\eta_{\bm{k}}|^2}{2}\sum_{i=1}^3\rho_i^2\right] \mathrm{Re}\left[ e^{-p_0} \right]
\nonumber  \\ & \times
	\left[ \cos\left(q_{\mathrm{LO}}v_{\bm{k},x}(\rho_1+\rho_2+\rho_3) \right)+\cos\left( q_{\mathrm{LO}}v_{\bm{k},x}(\rho_1+\rho_2-\rho_3) \right)
	+\cos\left(q_{\mathrm{LO}}v_{\bm{k},x}(\rho_1-\rho_2-\rho_3) \right) \right]
	\Bigg\rangle_{\hat{\bm{k}}} |\Delta|^4 ,
\end{align}
\begin{align}
	F_{\Delta}^{\mathrm{(6)}}=&- \frac{5\pi c_6 T N(0)}{\sqrt{3}} \int_0^\infty \prod_{i=1}^5 d\rho_i\ f\left( \sum_{i=1}^5\rho_i \right) 
	\left\langle |w_{\bm{k}}|^4 \exp\left[- \frac{|\eta_{\bm{k}}|^2}{2}\sum_{i=1}^5\rho_i^2 \right] \mathrm{Re}\left[e^{-p_1}\right]
	\right\rangle_{\hat{\bm{k}}} |\Delta|^6
\nonumber \\ & \quad
	+(\rho_2 \rightarrow \rho_2+\rho_3, \rho_3 \rightarrow -\rho_3, \rho_4 \rightarrow \rho_4+\rho_3) ,
\end{align}
\begin{align}
	F_{\Delta m}^{\mathrm{(2,2)}}=& N(0) \int_0^\infty \prod_{i=1}^3 d\rho_i\ \frac{4\pi T}{\sinh(2\pi T \sum_{i=1}^3 \rho_i)} \left\langle |w_{\bm{k}}|^2 [K^{\mathrm{(n)}}+K^{\mathrm{(an)}}]
	\right\rangle_{\hat{\bm{k}}} |\Delta|^2 |m|^2 , \label{eqApp:f22} 
\end{align}
where
\begin{align}
	K^{\mathrm{(n)}}=&
	 \left[
	\cos\left(2I(\rho_1+\rho_2) \right)\cos\left(\delta(\bm{k})\rho_3 \right) e^{ -|\eta_{\bm{k}}|^2(\rho_1+\rho_2)^2/2 }
	+\cos\left(2I\rho_2\right)\cos\left(\delta(\bm{k})(\rho_1+\rho_3) \right) e^{ -{|\eta_{\bm{k}}|^2\rho_2^2/2} } 
	\right]
\nonumber  \\ & \quad \times
	\left[ \cos\left(q_{\mathrm{LO}}v_{\bm{k},x}(\rho_1+\rho_2+\rho_3)\right)+\cos\left(q_{\mathrm{LO}}v_{\bm{k},x}(\rho_1-\rho_2+\rho_3)\right) 
	+\cos\left(q_{\mathrm{LO}}v_{\bm{k},x}(\rho_1+\rho_2-\rho_3)\right) \right] , 
	\label{eqApp:f22_n} \\  
	K^{\mathrm{(an)}}=&
	-\cos\left(2I(\rho_1-\rho_2)\right)\cos\left(\delta(\bm{k})\rho_3 \right) e^{ -|\eta_{\bm{k}}|^2(\rho_1-\rho_2)^2/2}  
	\left[ \cos\left(q_{\mathrm{LO}}v_{\bm{k},x}(\rho_1+\rho_2+\rho_3)\right)
\nonumber \right. \\ & \left. \quad
	+\cos\left(q_{\mathrm{LO}}v_{\bm{k},x}(\rho_1-\rho_2+\rho_3)\right)
	+\cos\left(q_{\mathrm{LO}}v_{\bm{k},x}(\rho_1-\rho_2-\rho_3)\right) 
	\right]	. \label{eqApp:f22_an} 
\end{align}
Here, $v_{\bm{k},x}$ is the $x$-component of $\bm{v_k}$ in the rotated coordinates, $\eta_{\bm{k}} = (\gamma^{-1/2}v_{\bm{k},y}-i\gamma^{1/2}v_{\bm{k},z})/(\sqrt{2}r_H) $, $c_4=1.67$, $c_6=2.59$, 
$p_0=\frac{1}{2}[{\eta_{\bm{k}}^*}^2(\rho_1^2+\rho_3^2)+\eta_{\bm{k}}^2\rho_2^2]-\frac{1}{4}[\eta_{\bm{k}}^*(\rho_1+\rho_3)-\eta_{\bm{k}}\rho_2]^2$,
and
\begin{align}
	p_1=& \left[-\frac{1}{2}\left({\eta_{\bm{k}}^*}^2\sum_{i\mathrm{:odd}}\rho_i^2+\eta_{\bm{k}}^2\sum_{i\mathrm{:even}}\rho_i^2 \right)
		+\frac{1}{6}\left(\eta_{\bm{k}}^*\sum_{i\mathrm{:odd}}\rho_i+\eta_{\bm{k}}\sum_{i\mathrm{:even}}\rho_i \right)^2
\nonumber \right.\\ &\left. \quad
	+\frac{1}{3}\left({\eta_{\bm{k}}^*}^2\sum_{(i,j)\mathrm{:odd}}(\rho_i-\rho_j)^2+\eta_{\bm{k}}^2\sum_{(i,j)\mathrm{:even}}(\rho_i-\rho_j)^2 \right) \right]_{\rho_6=0}  .
\end{align}
Changing the integration variables of eqs.(\ref{eqApp:f22}), (\ref{eqApp:f22_n}), and (\ref{eqApp:f22_an}) to $\rho=\rho_3$, $\Lambda=(\rho_1+\rho_2)/2$, and $\tau=\rho_1-\rho_2$,
we reach eqs.(\ref{eq:Fcl}) and (\ref{eq:K}) in the main text. 
\end{widetext}

\end{document}